\documentclass[11pt]{article}
\usepackage{moriond,epsfig}

\usepackage{graphicx}
\usepackage[intlimits]{amsmath}
\usepackage{amssymb}
\usepackage{xspace}
\usepackage{subfig}
\usepackage[figuresright]{rotating}

\newcommand{\ee}{\ensuremath{e^+e^-}\xspace}
\newcommand{\mumu}{\ensuremath{\mu^+\mu^-}\xspace}

\newcommand{\ccbar}{\ensuremath{c\overline{c}}\xspace}
\newcommand{\bbbar}{\ensuremath{b\overline{b}}\xspace}
\newcommand{\pion}{\ensuremath{\pi^0}\xspace}
\newcommand{\eexp}[1]{\ensuremath{{\rm e}^{#1}}\xspace}

\newcommand{\fig}[1]{{Fig.~\ref{#1}}\xspace}

\newcommand{\eg}{e.\,g.\xspace}%
\newcommand{\ie}{i.\,e.\xspace}%
\newcommand{\etal}{{\it et al.}\xspace}%

\newcommand{\pp}{{\ensuremath{p+p}}\xspace}

\newcommand{\AuAu}{\ensuremath{\rm{Au}+\rm{Au}}\xspace}

\newcommand{\sqrtsnn}{\ensuremath{\sqrt{s_{NN}}}\xspace}

\newcommand{\mt}{\ensuremath{m_T}\xspace}
\newcommand{\pt}{\ensuremath{p_T}\xspace}

\newcommand{\gevc}{GeV/\ensuremath{c}\xspace}

\newcommand{\mevcc}{MeV/\ensuremath{c^2}\xspace}
\newcommand{\gevcc}{GeV/\ensuremath{c^2}\xspace}

\newcommand{\mee}{\ensuremath{m_{ee}}\xspace}

\newcommand{\taa}{\ensuremath{T_{AA}}\xspace}

\newcommand{\pythia}{{\sc Pythia}\xspace}

\begin{document}
\vspace*{4cm}
\title{Measurement of the dielectron continuum in \pp and \AuAu
  collisions at RHIC}

\author{ T. DAHMS, for the PHENIX Collaboration }

\address{CERN,\\
  1211 Geneva 23, Switzerland}

\maketitle
\abstracts{PHENIX has measured the \ee pair continuum in \sqrtsnn =
  200 GeV \AuAu and \pp collisions over a wide range of mass and
  transverse momenta. While the \pp data in the mass range below the
  $\phi$ meson are well described by known contributions from light
  meson decays, the \AuAu minimum bias inclusive mass spectrum shows
  an enhancement by a factor of $4.7 \pm 0.4{\rm (stat)} \pm 1.5{\rm
    (syst)} \pm 0.9{\rm (model)}$ in the mass range
  $0.15<\mee<0.75$~\gevcc. At low mass ($\mee < 0.3$~\gevcc) and high
  \pt ($1<\pt<5$~\gevc) an enhanced \ee pair yield is observed that is
  in qualitative agreement with hydrodynamical models of thermal
  photon emission with initial temperatures ranging from $T_{init}
  \simeq$ 300--600~MeV at times of 0.6--0.15~fm/$c$ after the
  collision.}

\section{Introduction}
Experimental results from the Relativistic Heavy Ion Collider (RHIC)
have established the formation of dense partonic matter in \AuAu
collisions at \sqrtsnn = 200 GeV~\cite{whitepaper}. One of the key
questions has been what is the initial temperature of the created
matter. Like any medium in thermal equilibrium, the quark-gluon plasma
emits black-body radiation characteristic for its temperature in form
of real and virtual photons, the latter ones appearing as lepton pairs
(\ee or \mumu). As photons and lepton pairs do not carry color charge,
they do not undergo strong final state interaction while traversing
the medium. Thus, they carry all their information from the time they
have been created to the detector. But, because they are emitted
during all stages of the collision, any measurement will be a time
integrated average and the initial temperature can only be derived by
comparing to models. One of the main challenges of measuring thermal
photons is the huge background due to hadron decay photons. The
dominant background contribution due to \pion decays can be avoided by
measuring virtual photons rather than real photons. Any process that
creates a real photon can also create a virtual photon which decays
into a lepton pair. The invariant mass spectrum of virtual photons
created by the quark-gluon Compton scattering ($q+g \rightarrow
q+\gamma^{*}$) is described by the Kroll-Wada
equation~\cite{kroll-wada}. For direct virtual photons with
$\pt\gg\mee$ this spectrum is proportional to $1/\mee$. Also \ee pairs
from \pion Dalitz decays have the same $1/\mee$ shape at $\mee \ll
m_{\pion}$, but are suppressed due to the limited phase space when
approaching $m_{\pion}=135$~\mevcc. Thus, one can measure direct
virtual photons above the \pion mass where the main background source
is eliminated.

\section{Results}
\begin{sidewaysfigure}[p]
  \begin{center}
    \subfloat[]{\label{fig:mass_pp}\includegraphics[height=0.49\textheight]{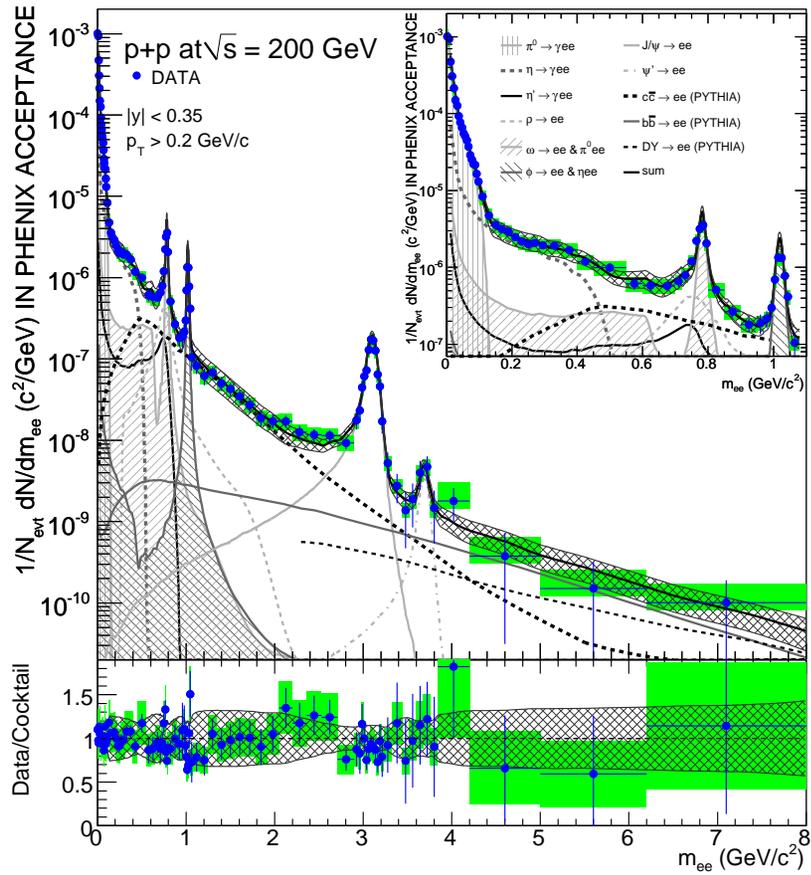}}
    \subfloat[]{\label{fig:mass_auau}\includegraphics[height=0.49\textheight]{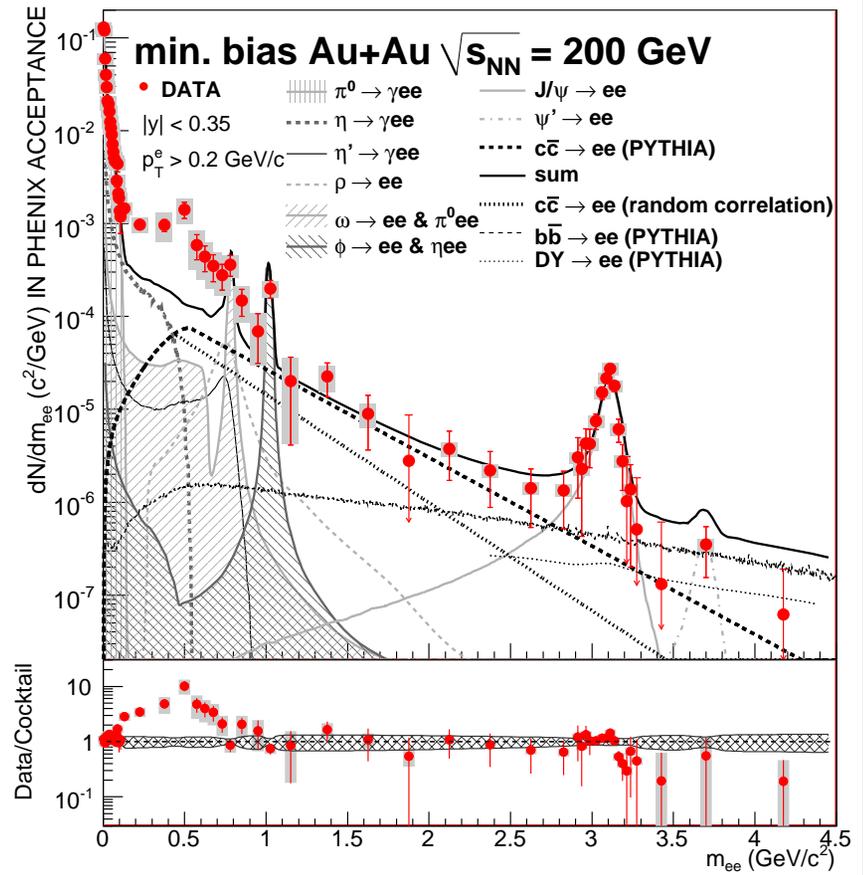}}
    \caption[]{Invariant mass distribution of \ee pairs in the PHENIX
      acceptance in \pp \subref{fig:mass_pp} and \AuAu
      \subref{fig:mass_auau} collisions. The data are compared to the
      expectations from the decays of light hadrons and correlated
      decays of open charm, bottom, and Drell-Yan.}
    \label{fig:invmass}
  \end{center}
\end{sidewaysfigure}

PHENIX has measured \ee pairs from \pp and \AuAu collisions at
\sqrtsnn = 200 GeV as a function of mass and \pt. The resulting
invariant mass spectra integrated over all \pt are shown in
\fig{fig:invmass}. Contributions from combinatorial and correlated
background sources have been subtracted statistically utilizing mixed
events and the like-sign pairs. The details of the analysis have been
described in Ref.~\cite{ppg085,ppg088}. The \pp data are compared to a
cocktail of the expected hadronic sources whose cross sections have
been measured independently by PHENIX~\cite{ppg088}. The contribution
from semi-leptonic charm and bottom decays as well as Drell-Yan have
been simulated with \pythia~\cite{pythia}. The agreement between data
and simulation over the full mass range is excellent. By comparing the
integrated yield of data and \pythia in the intermediate mass region
of $1.1 < \mee < 2.5$~\gevcc, which is dominated by open charm, a
total charm cross section of $\sigma_{\ccbar} = 544\pm39{\rm
  (stat)}\pm142{\rm (syst)}\pm200{\rm (model)}~\mu$b has been measured
in \pp collisions. Alternatively, by fitting the normalization of the
simulated shapes to the data, the charm and bottom cross sections of
$\sigma_{\ccbar} = 518\pm47{\rm (stat)}\pm135{\rm (syst)}\pm190{\rm
  (model)}~\mu$b and $\sigma_{\bbbar} = 3.9{\rm
  (stat)}\pm2.4^{+3}_{-2}{\rm (syst)}~\mu$b, respectively, have been
measured~\cite{ppg085}. The charm cross section is in agreement with
the result of the non-photon single electron
measurement~\cite{ppg065}: $\sigma_{\ccbar} = 567\pm57{\rm
  (stat)}\pm193{\rm (syst)}~\mu$b.

The measurement of \ee pairs in \AuAu collisions is compared to the
corresponding cocktail calculation in \fig{fig:mass_auau}. The charm
contribution is the same \pythia calculation scaled to the number of
binary collisions, $N_{\rm coll}$, times the measured charm cross
section of $\sigma_{\ccbar} = 567\pm57{\rm (stat)}\pm193{\rm
  (syst)}~\mu$b. In contrast to the \pp result, the measured yield of
low mass region ($0.15<\mee<0.75$~\gevcc) in minimum bias \AuAu is
enhanced compared to the expectation by a factor $4.7\pm0.4{\rm
  (stat)}\pm1.5{\rm (syst)}\pm0.9{\rm (model)}$. The intermediate mass
region is surprisingly well described by \pythia which is shown as
dashed line in \fig{fig:mass_auau}. This is interesting to note, as
single electron distributions from semi-leptonic charm decays show
substantial medium modifications~\cite{ppg066}. Thus, it is hard to
understand how the dynamical correlation of the \ccbar pair at
production remains unaffected by the medium. An alternative scenario
in which this correlation is washed out by the medium, \ie the
direction of the $c$ and the $\overline{c}$ quark are uncorrelated, is
indicated by a dotted line in the same figure. This leads to a much
softer spectrum and would leave significant room for other
contributions, \eg thermal radiation emited by quark-antiquark
annihilation processes.

\begin{sidewaysfigure}[p]
  \begin{center}
    \subfloat[]{\label{fig:mass_pt_pp}\includegraphics[width=0.44\textwidth]{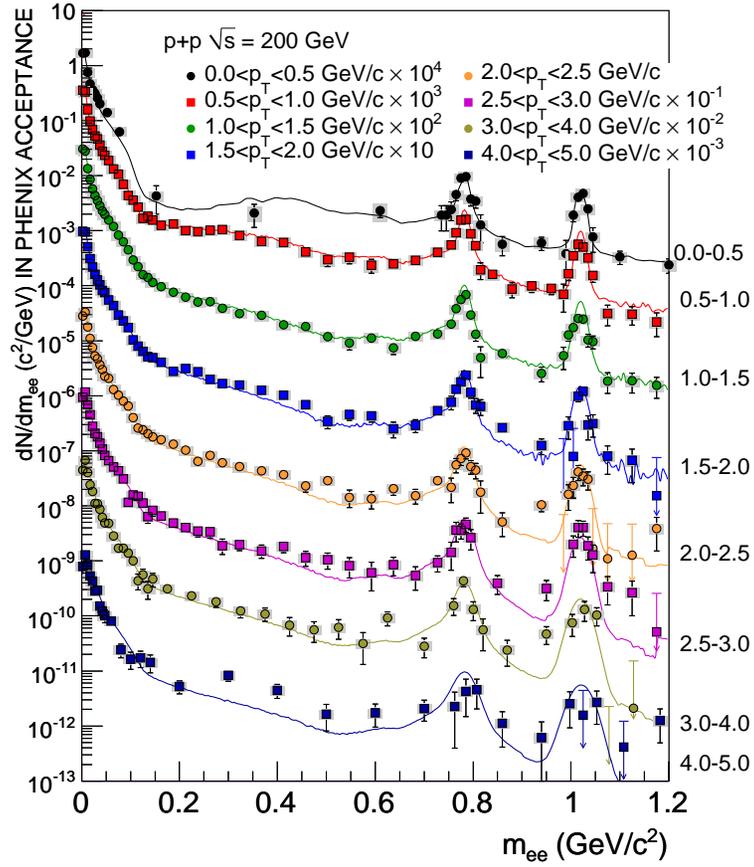}}
    \subfloat[]{\label{fig:mass_pt_auau}\includegraphics[width=0.44\textwidth]{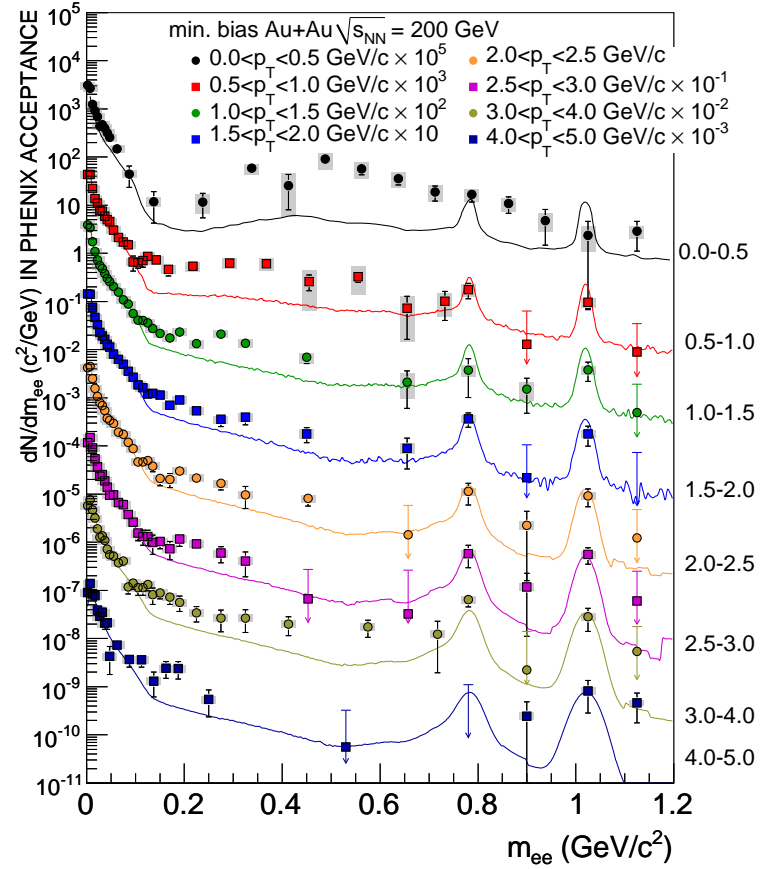}}
    \caption[]{Invariant mass distribution of \ee pairs in the PHENIX
      acceptance for different \pt ranges in \pp \subref{fig:mass_pp}
      and \AuAu \subref{fig:mass_auau} collisions. The data are
      compared to the cocktail of hadronic sources which includes the
      contribution from charm as described in the text.}
    \label{fig:mass_pt}
  \end{center}
\end{sidewaysfigure}

\begin{sidewaysfigure}[p]
  \begin{center}
    \subfloat[]{\label{fig:direct_xsec}\includegraphics[height=0.49\textheight]{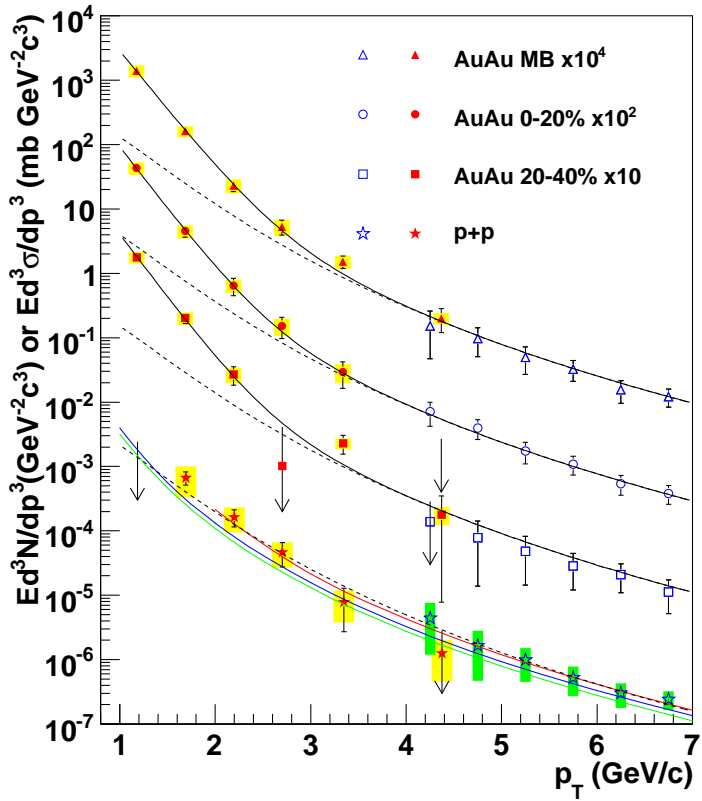}}
    \subfloat[]{\label{fig:direct_theory}\includegraphics[height=0.45\textheight]{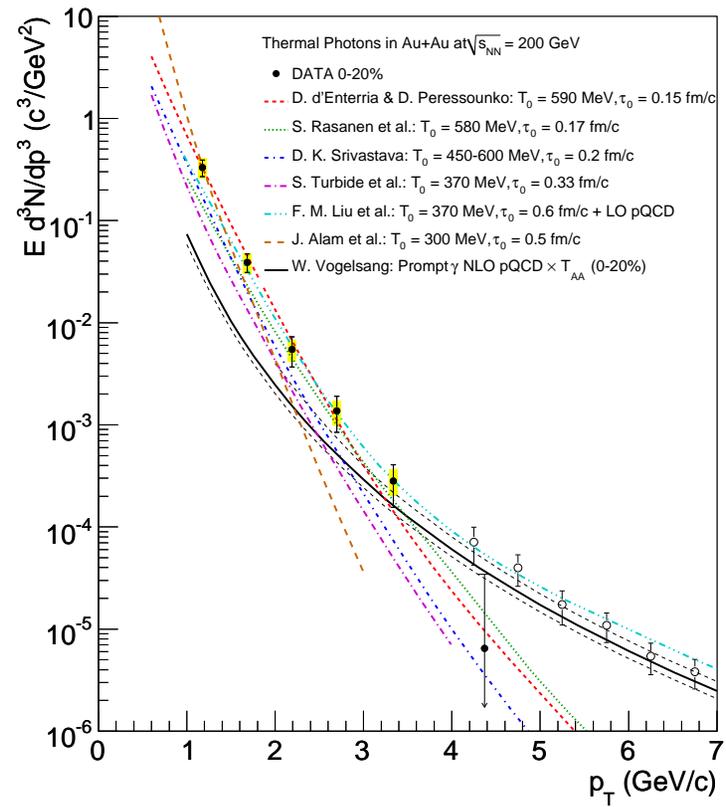}}
    \caption[]{The invariant cross section (\pp) and invariant yield
      (\AuAu) of direct photons as a function of \pt is shown in
      \subref{fig:direct_xsec}. Solid points are from the internal
      conversion analysis, open points from~\cite{ppg042,ppg060}. The
      \pp data are compared to NLO pQCD calculations~\cite{vogelsang}
      for three theory scales $\mu = 0.5\,\pt$, \pt, and $2\,\pt$. The
      \pp data are fit to a modified power-law (dashed curve) which is
      scaled by \taa and compared to the \AuAu data. The \AuAu data
      are fitted by the \taa scaled \pp fit plus an exponential (solid
      lines). In \subref{fig:direct_theory} hydrodynamical model
      calculations of thermal photon emission are compared with the
      direct photon data in the 20\% most central \AuAu collisions. In
      contrast to the others, the curve of F. M. Liu
      \etal~\cite{liu08} includes contributions from pQCD. The black
      solid curve show the \taa scaled pQCD
      calculation~\cite{vogelsang}.}
    \label{fig:directpt}
  \end{center}
\end{sidewaysfigure}

The \pt dependence of the low mass region ($\mee < 1.2$~\gevcc) is
shown in \fig{fig:mass_pt}. The \ee pairs measured in \pp are
consistent with the cocktail for the lower \pt bins. At higher \pt
however, the data are above the expectation. In \AuAu the enhancement
is concentrated at low \pt but extends to the high \pt region where it
is still significantly higher than in \pp. The shape of the low mass
enhancement in \AuAu differs for $\pt<1$~\gevc and $\pt>1$~\gevc. At
high \pt the excess can be interpreted as due to internal conversions
of direct virtual photons which, for $\pt\gg\mee$, has a well defined
$1/\mee$ mass dependence~\cite{kroll-wada}. The data have been fit to
the cocktail plus a direct photon contribution in the form of $f(\mee)
= (1-r) f_{c} (\mee) + r f_{\rm dir}(\mee)$ with the fraction of the
direct virtual photon contribution $r$ as the free parameter. From
this fraction of the direct virtual photon contribution it is possible
to derive the \pt spectrum of direct real photons~\cite{ppg086} which
is shown for \pp and three \AuAu centrality bins (min. bias, 0--20\%,
and 20--40\%) in \fig{fig:direct_xsec}. The data are extended at high
\pt with the previous measurements of direct photons in the
electromagnetic calorimeter of PHENIX~\cite{ppg042,ppg060} which are
in good agreement in the overlapping \pt range. The \pp data are are
in good agreement with a NLO pQCD calculation of direct
photons~\cite{vogelsang} which is shown for three different theory
scales: $\mu=0.5\,\pt$, \pt, and $2\,\pt$.

In order to compare the \pp data quantitatively with the \AuAu
measurement, the \pp direct photon spectrum is fit to a modified
power-law function $A_{pp} (1+\pt^2/b)^{-n}$.  This fit, scaled by the
nuclear overlap function \taa is overlaid with the \AuAu data. For all
three centrality bins the data are show a clear enhancement at low
\pt. This excess is characterized by fitting the \taa scaled \pp fit
plus an exponential $A\eexp{-\pt/T} + \taa \times A_{pp}
(1+\pt^2/b)^{-n}$.  For the 20\% most central collisions, the inverse
exponential slope is $T = 221\pm 19{\rm(stat)}\pm19{\rm (syst)}$.  The
result for this centrality bin is compared in \fig{fig:direct_theory}
to several hydrodynamical models of thermal photon
emission~\cite{dEnterria:2005vz,huovinen02,srivastava01,Turbide:2003si,liu08,alam01}.
The models, which assume the formation of a hot system with initial
temperatures ranging from $T_{\rm init} = 300$~MeV at a thermalization
time of $\tau_0 = 0.6$~fm/$c$ to $T_{\rm init} = 600$~MeV at $\tau_0 =
0.15$~fm/$c$, are all in qualitative agreement with the data. Lattice
QCD predicts a phase transition from the hadronic phase to the
quark-gluon plasma at $T\simeq170$~MeV.

While for $\pt>1$~\gevc the excess yield can be successfully described
by internal conversions of direct virtual photons, the excess yield at
lower \pt, which is responsible for most of the \pt integrated
enhancement, shows a different mass dependence. The transverse mass
dependence of the enhancement has been analysed in
Ref.~\cite{ppg088}. While not shown in this proceedings, it appears
that, in addition to the thermal radiation at high \mt, there is a
second much softer component at low \mt with an inverse slope of
$T\sim90$~MeV. This low mass, low \pt enhancement is currently not
described by any theoretical model calculation.

\end{document}